%% file: paper-v1.tex
\documentclass[11pt, a4paper, twocolumn]{article} 

\input{structure.tex}

\usepackage{subfig}
\usepackage{hyperref}

\title{Revisiting the Arguments for\\Edge Computing Research}

\author{
	\authorstyle{
	Blesson Varghese\textsuperscript{1},
    Eyal de Lara\textsuperscript{2}, 
    Aaron Ding\textsuperscript{3},
    Cheol-Ho Hong\textsuperscript{4},
    Flavio Bonomi\textsuperscript{5},
    Schahram Dustdar\textsuperscript{6},
    Paul Harvey\textsuperscript{7},
    Peter Hewkin\textsuperscript{8},
    Weisong Shi\textsuperscript{9},
    Mark Thiele\textsuperscript{8},
    Peter Willis\textsuperscript{10}}
    \newline\newline
	\textsuperscript{1}\institution{Queen's University Belfast, UK }
	\textsuperscript{2}\institution{University of Toronto, Canada }
	\textsuperscript{3}\institution{TU Delft, Netherlands }
	\textsuperscript{4}\institution{Chung-Ang University, S. Korea }
	\textsuperscript{5}\institution{Lynx Software Technologies, USA }
	\textsuperscript{6}\institution{TU Wien, Austria }
	\textsuperscript{7}\institution{Rakuten Mobile, Japan }
	\textsuperscript{8}\institution{SmartEdge Datacentres Ltd., UK/USA }
	\textsuperscript{9}\institution{Wayne State University, USA }
	\textsuperscript{10}\institution{British Telecommunications plc, UK }
} 
	
\begin{document}

\date{\it Preprint Version - June 2021}

\maketitle 
\thispagestyle{firstpage} 

\input{abstract}

\section*{}
\input{introduction}
\label{sec:introduction}

\section*{Latency}
\input{latency}

\section*{Bandwidth}
\input{bandwidth}

\section*{Proliferation}
\input{proliferation}

\section*{Sustainability}
\input{sustainability}

\section*{Privacy and Sovereignty}
\input{privacy-sovereignty}

\section*{Conclusion}
\input{conclusion}

\section*{Acknowledgment}
\input{acknowledgment}

\noindent \textbf{Blesson Varghese} is Associate Professor of Computer Science at Queen's University Belfast, UK, and a Royal Society Short Industry Fellow. Contact him at b.varghese@qub.ac.uk. \\

\noindent \textbf{Eyal de Lara} is Full Professor of Computer Science at the University of Toronto, Canada. Contact him at delara@cs.toronto.edu.\\

\vfill\pagebreak

\noindent \textbf{Aaron Ding} is Assistant Professor at the TU Delft, Netherlands, and Adjunct Professor at the University of Helsinki, Finland. Contact him at aaron.ding@tudelft.nl. \\

\noindent \textbf{Cheol-Ho Hong} is Associate Professor of Electrical and Electronics Engineering at the Chung-Ang University, Korea. Contact him at cheolhohong@cau.ac.kr. \\

\noindent \textbf{Flavio Bonomi} is Technology Advisor to the Board of Directors, Lynx Software Technologies, USA. Contact him at fgbonomi@gmail.com. \\

\noindent \textbf{Schahram Dustdar} is Fellow of IEEE and a Full Professor of Computer Science at the TU Wien, Austria. Contact him at dustdar@dsg.tuwien.ac.at.\\ 

\noindent \textbf{Paul Harvey} is Research Lead at the Autonomous Networking Research and Innovation Department, Rakuten Mobile, Japan. Contact him at paul.harvey@rakuten.com. \\ 

\noindent \textbf{Peter Hewkin} is Founder of SmartEdge Data Centres Ltd., UK. Contact him at ph@smartedgedc.com. \\


\noindent \textbf{Weisong Shi} is Fellow of IEEE, a Charles H. Gershenson Distinguished Faculty Fellow and a Full Professor of Computer Science at the Wayne State University, USA. Contact him at weisong@wayne.edu. \\

\noindent \textbf{Mark Thiele} is Director of the SmartEdge Data Centres Ltd., USA, and CEO of Edgevana. Contact him at mt@smartedgedc.com. \\

\noindent \textbf{Peter Willis} is Chief Data Networks Strategist at the Research and Innovation Division, British Telecommunications plc, UK. Contact him at peter.j.willis@bt.com. \\

\end{document}

%% file: structure.tex
\usepackage[english]{babel} 
\usepackage{pdfcomment}

\usepackage{microtype} 

\usepackage{amsmath,amsfonts,amsthm} 

\usepackage[svgnames]{xcolor} 

\usepackage[hang, small, labelfont=bf, up, textfont=it]{caption} 

\usepackage{booktabs} 

\usepackage{lastpage} 

\usepackage{graphicx} 

\usepackage{enumitem} 
\setlist{noitemsep} 

\usepackage{sectsty} 
\allsectionsfont{\usefont{OT1}{phv}{b}{n}} 

\usepackage{geometry} 

\usepackage[T1]{fontenc} 
\usepackage[utf8]{inputenc} 

\usepackage{XCharter} 

\usepackage{fancyhdr} 
\fancypagestyle{firstpage}{ 
	\fancyhf{}
}

\newcommand{\authorstyle}[1]{{\large\usefont{OT1}{phv}{b}{n}\color{DarkRed}#1}} 

\newcommand{\institution}[1]{{\footnotesize\usefont{OT1}{phv}{m}{sl}\color{Black}#1}} 

\usepackage{titling} 

\newcommand{\HorRule}{\color{DarkGoldenrod}\rule{\linewidth}{1pt}} 

\pretitle{
	\vspace{-30pt} 
	\HorRule\vspace{5pt} 
	\fontsize{22}{26}\usefont{OT1}{phv}{b}{n}\selectfont 
	\color{DarkRed} 
}

\posttitle{\par\vskip 10pt} 

\preauthor{} 

\postauthor{ 
	\vspace{5pt} 
	\par\HorRule 
	\vspace{5pt} 
}

\usepackage{lettrine} 
\usepackage{fix-cm}	

\newcommand{\initial}[1]{ 
	\lettrine[lines=3,findent=4pt,nindent=0pt]{
		\color{DarkGoldenrod}
		{#1}
	}{}%
}

\usepackage{xstring} 

\newcommand{\lettrineabstract}[1]{
	\StrLeft{#1}{1}[\firstletter] 
	\initial{\firstletter}\textbf{\StrGobbleLeft{#1}{1}} 
}

\usepackage{cite}

\usepackage{array}
\newcolumntype{L}[1]{>{\raggedright\let\newline\\\arraybackslash\hspace{0pt}}m{#1}}
\newcolumntype{C}[1]{>{\centering\let\newline\\\arraybackslash\hspace{0pt}}m{#1}}
\newcolumntype{R}[1]{>{\raggedleft\let\newline\\\arraybackslash\hspace{0pt}}m{#1}}

%% file: abstract.tex
\lettrineabstract{This article argues that low latency, high bandwidth, device proliferation, sustainable digital infrastructure, and data privacy and sovereignty continue to motivate the need for edge computing research even though its initial concepts were formulated more than a decade ago.
}

%% file: introduction.tex
The initial concepts of edge computing were formulated more than a decade ago~\cite{cloudlet-01}. 
Although a nascent research area, it is generally understood that edge computing enables the (pre)processing of data closer to the source outside a centralized and geographically distant data center. Although not articulated in its current form, there were several notions of geography-aware computing in previous decades with a premise to bring compute services closer to data. 

`Edge' generally refers to a location rather than any specific technology for computing. However, it has started to emerge that the edge may need to be more than a location. 
Recent advances in 5G, AI and processor technologies and their application in novel domains have necessitated a strong need for geography-aware computing. Thus, edge computing has received attention which has inadvertently coupled the notion of the edge as a location with certain technologies.

An exemplar of edge computing that is commercially used is Content Delivery Networks (CDNs). 
They are commonly used to deliver digital content (web, gaming, AR/VR, videos) from servers to end-users by Internet Service Providers, carriers and network operators. 
More than half of today’s consumer traffic is generated in delivering digital content to users in the Internet using CDNs. 
Digital content is replicated and stored across many edge servers in different geographic locations, a concept referred to as ‘edge caching’, which is commercially used for improving application responsiveness and reducing latencies. 

When the cloud was rapidly being adopted within the technology landscape, it was argued that extremely centralized compute resources of the cloud would not be suitable for a wide-range of sensor-rich applications that were to emerge in the future. 
End-user devices or sensors generate data in these applications that is transferred elsewhere for processing (as opposed to delivering content from servers to end-users). 
Such applications would be latency-critical, bandwidth-intensive, and privacy-craving. 
A few hyperscalers and comparatively low network speeds observed then mandated the need for more decentralized data centers to be placed and used at the edge.  
However, it was always recognized that hyperscalers as economies at scale were essential and could not become redundant infrastructure. 

Times have now changed - there are plenty of cloud data center locations scattered across the globe and data can travel through advanced fiber optic communication channels at (near) speed of light. Do the arguments that initially mandated the need for edge computing still hold? 

Recent research articles examined cloud reachability across the globe to measure the average round-trip communication latency for an end-user when communicating with the cloud~\cite{reachability-01, reachability-02}. The authors concluded that current clouds in the United States were sufficient for many latency-critical applications and noted that the motivation for realizing edge computing as a mere \textit{`enthusiasm for newer computing paradigms'} (the data used in the above mentioned research and the conclusions will be examined in the next section).  

Contrary to the above, we note that cloud and edge computing are not necessarily competing paradigms; rather they are compatriots in delivering computing as a ubiquitous utility by appealing to arguments that will be discussed in this article. 
In light of the above and a renewed interest in determining whether there is still a need for edge computing both as a concept and an avenue of research, this article (re)examines five different arguments, namely (1) Latency, (2) Bandwidth, (3) Proliferation, (4) Sustainability, and (5) Privacy and Sovereignty. 

%% file: latency.tex
Reducing the overall latency in processing data at the source or delivering data from servers to end-users has been a key argument in favor of edge computing. 
These arguments have been supported by predictions of Gartner, for example, anticipating that by 2025, over 50\% of enterprise data will be created and processed outside the typical data center (\url{https://gtnr.it/3wzgTpf}).

We note that different technology providers consider latency in diverse ways. Therefore, some clarity is required on what should constitute the latency metric. For example, consider an end-user connected via a wired broadband connection - latency should refer to the sum of the times for raising a request from the source (for example, a device), for transporting the request over the network (including the delays incurred on different hops), for processing the request on the receiving server, for sending the response back to the source, and for taking an action on the source. The transport time from the source to the server and back only accounts for the round-trip communication latency. Often computational latencies are ignored. 
When considering a mobile network, the round trip latency between the source and the access network should also be accounted for. 

The Federal Communications Commission of the United States (US) carried out a performance measurement study of broadband services in the US. Ten major Internet Service Providers (ISPs) and an additional nine organizations participated in the exercise (\url{https://bit.ly/3gyptPx}).
The measurement servers were located in thirteen cities across the US with multiple locations in each city. The median round trip communication latencies observed on fiber optic cables were between 10~ms to 27~ms to edge locations. 

Broadband connection latencies enable us to quantify what delays will be incurred within an enclosed environment, such as a home or office. Given that a vast number of users rely on mobile devices and that machine-to-machine, vehicle-to-vehicle, machine-to-everything and vehicle-to-everything will need to rely on telecommunication infrastructure, it is worthwhile considering mobile network latencies. 4G, which is the most available global mobile network model has observed communication latencies of over 30~ms. In 2019, Opensignal reported that only 13 countries had a communication latency of between 30-40~ms which excluded North America and many parts of Europe (\url{https://bit.ly/2TGEvJY}).
These reported latencies are access network latencies and do not include the latency for reaching an edge compute location via the mobile network or for performing computations. With 5G, although a theoretical 1~ms communication latency is envisioned, early deployments in the US in 2019 had demonstrated nearly a 30~ms communication latency for the access network. In the UK, the 5G deployments in 2020 had a communication latency of at least 20~ms for the access network. 

The above communication latencies can indeed support many interactive applications that are already in use today. However, they will not be adequate to support (near) real-time computing (sub millisecond), such as those required for rapid responsiveness of autonomous cars or robots. For these contexts, the overall latency will need to be guaranteed. Hence, it would not be sufficient for any latency measuring exercise to merely highlight the average of a distribution of observed communication latencies without considering computation latencies and the type of application. In addition, the tail-end and outlier latencies in a distribution may be substantially higher than the average latency which also need to be accounted for. 

At this point, the dataset employed by the research articles investigating cloud reachability is considered~\cite{reachability-01, reachability-02}.
The dataset employed is from RIPE Atlas, an Internet measurement network that provides hardware probes for network measurements (for example, ping) that is publicly available (\url{https://mediatum.ub.tum.de/1593899}).
We note that these measurements reflect only network communication latencies and do not include computation latencies associated with the execution of application code.

\begin{figure*}
    \centering
    \subfloat[Boxplot of probe location latencies shorter in order of increasing average latency. Whiskers show minimum and 99\% latency.]{
        \label{fig:latency_boxplot}
        \includegraphics[width=\columnwidth]{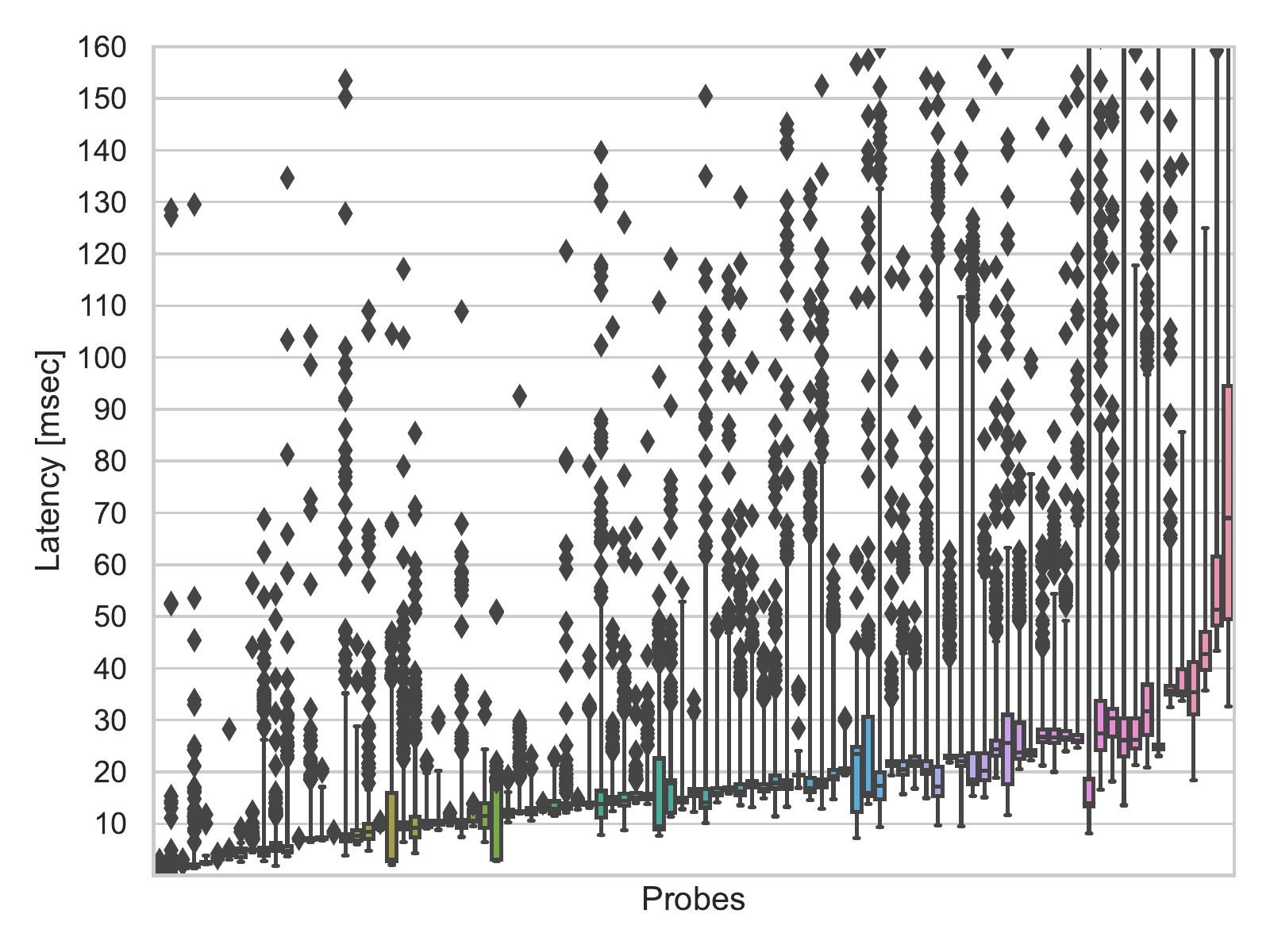} 
        }
    \subfloat[CDF of latencies for probe locations.]{
        \label{fig:latency_cdf}
        \includegraphics[width=\columnwidth]{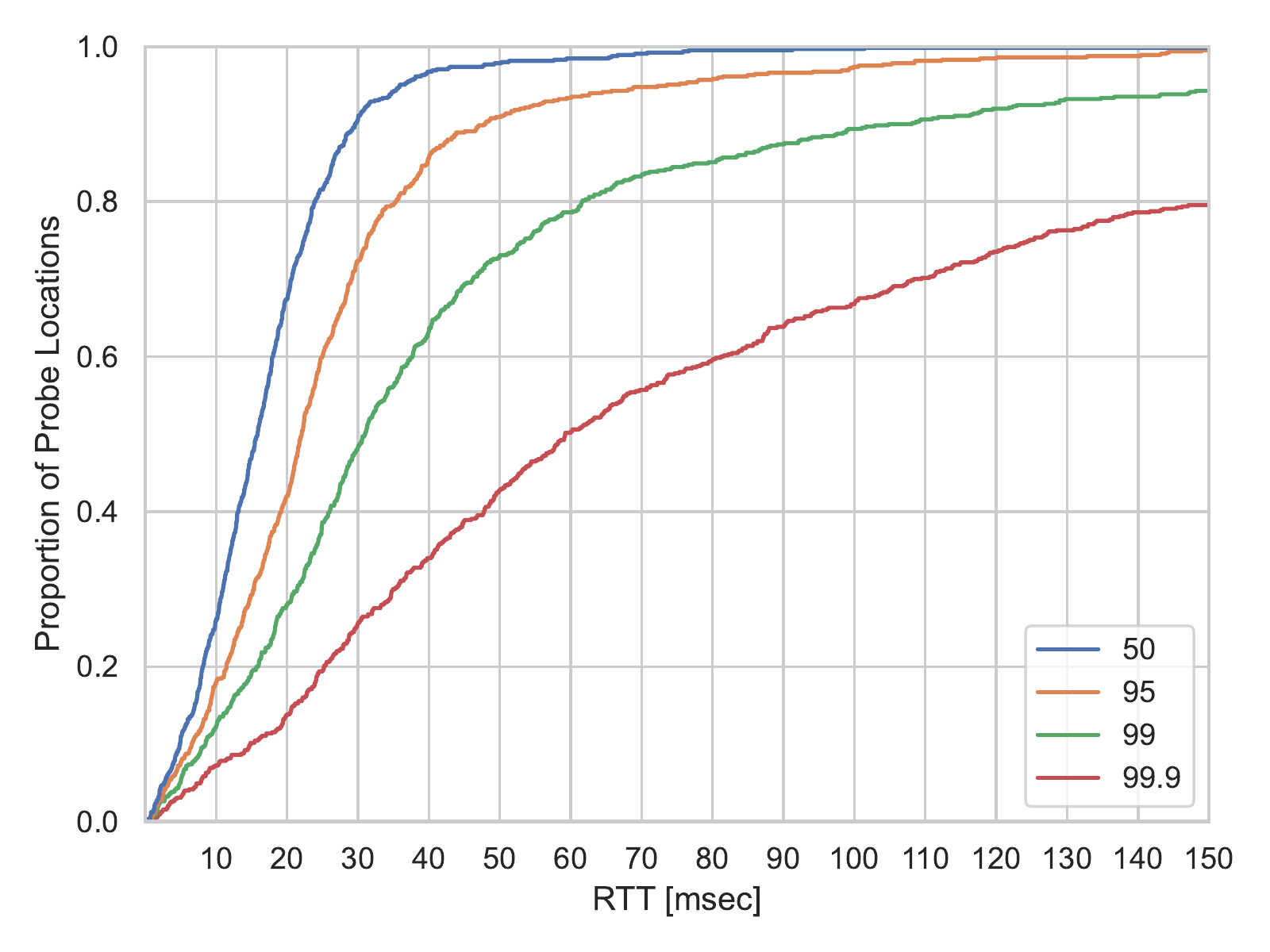}
        }
    \caption{Latency to closest cloud data center for probe locations in the USA with at least 100 measurements.}
    \label{fig:latency}
\end{figure*}

We analyzed the dataset and focused on the data for the United States containing 3091 different probe locations. 
For each location, there are measurements for up to 102 different data centers.
Only the closest data center for each probe as determined by the lowest average latency was considered.
Since 80\% of the locations have less than 64 measurements per data center, we focused on the remaining 650 locations that have at least 100 measurements to their closest data center; the average no. of measurements per probe is 2611.

Figure~\ref{fig:latency} shows the results of our analysis.  Figure~\ref{fig:latency_boxplot} shows a box plot of the latency distributions sorted by increasing average latency. For clarity, the plot only includes 1 out of every 7 probes (the plot of the complete dataset shows a similar pattern but is is very hard to read due to clutter). The top and bottom of the box represent 25\% and 75\% latencies, and the whiskers show the minimum and 99\% latency. Measurements outside of this range are shown as individual outliers. Figure~\ref{fig:latency_cdf} shows the cumulative distribution for the proportion of probes which experience median, 95\%, 99\%, and 99.9\% latency below a given threshold. For example, the figure shows that 25.4\% of probe locations experience a median latency to their closest data center of 10~milliseconds or less.

We observed that the majority of locations had a round-trip communication latency of more than 10~milliseconds. Moreover, even probe locations that experience low median latency observe very substantial variations. For example, only 6.7\% of the 650 locations were able to reach their closest data center within 10~milliseconds 99.9\% of the time. This rose to 18\% of the locations when lowering to 95\% of the time.  

The current communication latencies observed to the nearest cloud locations are undoubtedly an improvement over the average of 80~ms that were observed when edge computing was initially formulated as a concept~\cite{oldlatency-01}. 
Overall latencies under 10~milliseconds (let alone sub-milliseconds) cannot be guaranteed today on current public clouds for applications that require performance guarantees. 
Latency measurement studies are required to better understand edge computing. However, focusing on average latency~\cite{reachability-01, reachability-02} does not paint a correct or complete picture as it inherently hides significant variations in network latency over time.  

The above have led to new industry trends that will potentially lead to the convergence of what is today known as the cloud and edge. For example, cloud providers are embracing edge locations for setting up data centers on the last mile network (for example, Amazon Outpost) together with dedicated hardware, such as the AZ1 neural edge processors for the extreme edge to reduce communication latencies. 

However, edge as a location is only one aspect of the latency argument.
If only communication latencies had to be considered, then edge compute locations would need to be placed every 60 miles for theoretically achieving a 1~ms round trip communication latency using current fibre optic technologies (based on the speed of light in a medium with a refractive index of 1.5) between two endpoints ignoring latencies in the access network, processing delays on the hops, network congestion or computation latencies on servers. Telecom providers are experimenting with hollow core fibre optics to transmit data at (near) speed of light for reducing latencies (\url{https://bit.ly/3cOH97l}).
The invasiveness and substantial increase in costs of infrastructure may not be pragmatic for a global rollout and that by reducing communication latencies alone may not be sufficient for minimizing the overall latency. 

There are a select few locations around the globe by virtue of geographic location or proximity to traditional data centers that can achieve an average communication latency of 10~ms. Nonetheless, delivering low overall latency globally for emerging and futuristic applications still remains a challenge to be surmounted and a vision to be fulfilled. Transformative advancements are still required both on the networking and computing fronts to achieve this. Thus, latency continues to be a first-class argument for edge computing research. 

%% file: bandwidth.tex
The network bandwidth bottleneck of the wide-area network (WAN) to the cloud has been another argument in favor of edge computing~\cite{edge-02}. It was demonstrated that the network bandwidth on a WAN is restrictive due to the number of traversed hops ranging from 9 to 20~\cite{persico2017performance}. The bandwidth between two Amazon EC2 virtual machines (VMs) in the same data center was 900~Mbps in 2015~\cite{persico2015measuring}. However, when the WAN is involved, the bandwidth to the same VMs was 30--160~Mbps~\cite{persico2017performance}. Furthermore, most cloud providers throttle the bandwidth when the total data transfer reaches a threshold. Therefore, a distant cloud is not adequate for emerging applications that require high network throughput.

Emerging applications including AR/VR, remote-controlled factories, and autonomous vehicles employ a wide range of devices and sensors at the edge of the network and increasingly generate (and consume) a large volume of data. Therefore, a high network bandwidth is required for meeting Quality-of-Service (QoS) objectives. Consider the example of autonomous vehicles. The Automotive Edge Computing Consortium (AECC) estimates that more than 30\% of video data produced on the vehicle will need to be offloaded. This is to increase safety thresholds by processing offloaded data with external data for augmenting awareness of the moving vehicle. The volume of data that will need to be offloaded is expected to be between 400 GB to 5 TB per hour. If all the data is sent to the cloud, the QoS objectives cannot not be met due to the limited bandwidth. Therefore, exploiting the edge that efficiently processes the data near the source is required for such applications.

Many devices and sensors are connected to the edge using the mobile network. The latest commercial 5G cellular network implements the millimeter wave (mmWave) technology, which theoretically offers bandwidth up to 20 Gbps for download and 10 Gbps for upload.
However, recent measurement studies in field tests of 5G mmWave performance in three major U.S. cities observed download speeds from 600 Mbps to 1.7 Gbps and upload speeds between 30 and 60 Mbps~\cite{narayanan2020first}. 
Similar download speeds and over three times higher upload speeds were observed on commercial 5G in China.
Since 5G has only begun commercialization, its performance is still far from the theoretical speed but offers higher bandwidth than 4G LTE. 

The current peak download speed of 5G mmWave is acceptable for many existing applications including video streaming and gaming. For example, high resolution cameras in a stadium can transmit a video stream directly to an edge server without sending the data to the cloud. The edge server then routes the stream to mobile devices in the same venue in order to avoid a latency delay. As the bandwidth required for 8K video streaming is 300 Mbps (\url{https://bit.ly/3zyes8i}), the current bandwidth of 5G can sufficiently support this application scenario. An emerging real-time streaming application such as volumetric videos, which capture three-dimensional space, demands throughput of at least 1.1 Gbps. The peak speed of current 5G can satisfy such a requirement, and the advances in 5G will be able to support more high quality volumetric videos in future. 


The current upload speed of 5G can meet the bandwidth requirements of non-bandwidth-hungry applications in edge computing. For example, 4K panoramic video telephony does not exceed the 5G upload capacity when sending all HD resolution videos up to 5.7K whereas 4G cannot support this. The uploaded video data can be processed at the edge in order to reduce the data volume, which will be transferred to users in different locations. This efficient data processing can provide low latency communication without exploiting the cloud. 

The current upload and download bandwidths available in 5G and to public clouds can satisfy the requirements of many existing applications. However, bandwidth is still a limiting factor that hinders the emergence of certain applications and continues to be an argument that motivates edge computing research. 

%% file: proliferation.tex
It is estimated that by 2025 more than 55 billion devices, sensors and instruments will be connected (\url{https://bit.ly/3q54VkI}).
This anticipated increase will consequently expose a larger attack surface. 
One key challenge is cybersecurity - detecting malicious users and containing breaches.

Detecting malicious activity is usually a data-driven approach and using extremely centralized resources to monitor are known to be challenging. Preceding versions of distributed computing paradigms have taught us that centralized monitoring is generally not scalable. Therefore, more distributed and hierarchical monitoring strategies are required which can find home on the edge~\cite{dimon}. In addition, intrusion detection and prevention systems, such as those used in vehicular ad-hoc networks are latency sensitive and the edge of the network is considered to be an ideal location~\cite{idps-01}. 

The edge appeals to providing more distributed locations for monitoring and data aggregation thereby inherently providing containment zones. Recent years have seen an increasing number of botnet and malware based attacks originating from IoT devices. Edge computing offers the opportunity for localized detection and isolation of such devices~\cite{botnet-01}. Network segmentation for example is one approach that can be adopted at the edge to contain the access of a potentially malicious device beyond the edge. 

Many existing edge applications only achieve a functionality improvement by using the edge - they may meet satisfactory performance thresholds even if what is known today as the cloud is available to them. However, looking forward, as edge-native workloads start to emerge, running services on the edge will eventually become necessities for people, factories, cities, and transportation that use them. Thus, even if networks beyond the edge were to fail, the edge can independently operate, thereby making our people and infrastructure more resilient.  

In relation to the device proliferation argument, edge computing is likely to pave way for achieving scalable decentralized management of security, enabling effective containment zones to isolate malicious devices, and delivering network independence for more resilience.

%% file: sustainability.tex
Sustainability may be understood in terms of electricity consumption, the amount of electricity to transmit data, and the consequent carbon footprint. 
The arguments on sustainability in complete favor of edge computing are not sufficiently well articulated and sometimes also send a mixed message. 
For example, on one hand Nature (\url{https://www.nature.com/articles/d41586-018-06610-y}) reported that it is anticipated by 2030 that nearly 21\% (other estimates say at least 8\%) of the worlds electricity consumption will be driven by increase in networks, requiring nearly 5,000 terawatt hours (TWh) per year and increase in data centers, requiring nearly 3,000~TWh per year~\cite{electricity2030-1}. 
The estimates presented assumed an exponential increase due to the expanding telecoms infrastructure and massively increasing internet traffic to and from data centers generated by end user devices/sensors and emerging applications. 


On the other hand, the IEA reported that the global data center energy demand has remained largely flat for the last ten years and data transmission networks have become more energy efficient (\url{https://bit.ly/2S5b6Jf}). 


There have been attempts to estimate the kilowatt hour per gigabyte of data (KWh/GB) transferred over the internet, but has resulted in values ranging across different orders of magnitude~\cite{electricitykwhgb-1}. All of the above suggests room for more large-scale measurement studies on further articulating the sustainability arguments. 

Nonetheless, it is commonly understood that there are costs involved in sending data over the networks. 
The energy required for transmitting data over the networks is at the least directly proportional to the distance that data needs to travel. With increasing data traffic it is only logical to consider localized data processing to reduce the overall amount of energy required by the networks. 
The data flowing through the internet is a primary driver for CO\textsubscript{2} emissions; other sources include from the Radio Access Network (RAN) and servers~\cite{c02-footprint-01}. By computing on the edge in a 5G network it was noted that the CO\textsubscript{2} footprint could be reduced by up to 50\%. 

Sustainability is therefore an important argument supporting edge computing research both from an electricity consumption and carbon footprint point-of-view, which are major global concerns. 
Data centers and networks indeed consume a large amount of electricity, but whether edge computing can substantially shift this trend is not yet clear.
Further insight from large-scale measurement exercises are required to make a more informed case. 

%% file: privacy-sovereignty.tex
Undoubtedly, data has become the fuel for the digital economy. Social welfare and advancement now relies on protecting critical data. Creating a trusted environment for all stakeholders (for example, public sector organizations, private organizations, governments and individual citizens) is underpinned by data privacy and sovereignty. 

There are significant privacy concerns as connected devices become data producers; large-scale machine learning in the cloud using data that is crowd-sourced from individual users may contain private information~\cite{Yu-SP2019}.
The edge is understood to meet this privacy gap by providing the unique capability of enforcing localized privacy control and establishing a trust proxy.

By leveraging the resource-rich layer between devices that generate data and distant clouds, the edge has been demonstrated in the context of distributed machine learning (such as federated learning) to achieve differential privacy for devices while meeting the regulatory and legislative requirements of data sovereignty, such as the General Data Protection Regulation (GDPR). This also aligns with the demand for data sovereignty in Canada, New Zealand, Australia and USA.
A more secure and trusted way of using personal data on the user edge has been demonstrated through the ‘Data Box’ approach.

The edge can better utilize local contexts practically to strike a balance between privacy and usability. Recent studies reveal the synergistic potential of edge, advanced machine learning and privacy-enhancing mechanisms~\cite{Gursoy-TDSC2019, lockhart2020scission}. 

The edge as an enabler for data privacy and sovereignty is an argument that will be further developed as the Internet is transformed into a more ethical system. Early research on privacy and sovereignty enhanced by the edge is encouraging. Therefore more collaborative efforts with researchers from disciplines outside the immediate technical envelope of edge computing (law, ethics and public policy) are required.

%% file: conclusion.tex
There are several arguments both technical and non-technical that continue to motivate edge computing research and innovation. The democratization of the future internet is yet another argument in favor of the edge~\cite{democratization-01}. The edge introduces new stakeholders (for example, providers, applications and users), enables the convergence of different technologies that have traditionally operated in silos and takes monopoly away from a select few global players and countries. As a part of this endeavor, the initiative on the federated data infrastructure for Europe GAIA-X (\url{https://bit.ly/3xu6o6N}) and the concept of the Global Data Plane~\cite{globaldataplane-01} recognize the edge as an essential building block for delivering open, transparent and trustworthy digital infrastructure. 


This article argues that the motivation for edge computing research has not diminished since it was first formulated.  
Ongoing edge research and the wide range of edge-native and edge-accelerated applications that are emerging are indications of the benefits of using the edge. Edge computing as an enabler for advancing new frontiers in space-based systems by reducing communication times and energy is one example among many (\url{https://ibm.co/3gA3Xdu}). 
While the case for edge computing in private networks and applications is clear and is now starting to become available to business customers (\url{https://reut.rs/3zAFb4k}), the value in a global public rollout awaits to be more precisely calculated. 


%% file: acknowledgment.tex

The feedback provided by numerous academics and industry experts during consultation is acknowledged. The first author is supported by a Royal Society Short Industry Fellowship.

%% file: paper-v1.bbl
\begin{thebibliography}{00}

\bibitem{cloudlet-01}
M. Satyanarayanan, P. Bahl, R. Caceres and N. Davies, ``The Case for VM-based Cloudlets in Mobile Computing,'' {\it IEEE Pervasive Computing}, vol. 8, no. 4, pp. 14--23, 2009. (journal)

\bibitem{reachability-01}
N. Mohan, L. Corneo, A. Zavodovski, S. Bayhan, W. Wong and J. Kangasharju, ``Pruning Edge Research with Latency Shears,'' {\it Proc. of the 19th ACM Workshop on Hot Topics in Networks}, pp. 182--189, 2020. (conference proceeedings)

\bibitem{reachability-02}
L. Corneo, M. Eder, N. Mohan, A. Zavodovski, S. Bayhan, W. Wong, P. Gunningberg, J. Kangasharju and J. Ott, ``Surrounded by the Clouds: a Comprehensive Cloud Reachability Study,'' {\it Proc. of the ACM Web Conference}, 2021. (conference proceedings)


\bibitem{oldlatency-01}
A. Li, X. Yang, S. Kandula and M. Zhang, ``CloudCmp: Comparing Public Cloud Providers,'' {\it Proc. of the 10th ACM SIGCOMM Conference on Internet Measurement}, pp. 1--14, 2010. (conference proceedings) 


\bibitem{edge-02}
W. Shi, J. Cao, Q. Zhang, Y. Li and L. Xu, ``Edge Computing: Vision and Challenges,'' {\it IEEE Internet of Things Journal}, vol. 3, no. 5, pp. 637--646, 2016. (journal)

\bibitem{persico2017performance}
V. Persico, A. Botta, P. Marchetta, A. Montieri and A. Pescap{\'e}, ``On the Performance of the Wide Area Networks Interconnecting Public Cloud Datacenters Around the Globe,'' {\it Computer Networks}, vol. 112, pp. 67--83, 2017. (journal)

\bibitem{persico2015measuring}
V. Persico, P. Marchetta, A. Botta and A. Pescap{\'e}, ``Measuring Network Throughput in the Cloud: The Case of Amazon EC2,'' {\it Computer Networks}, vol. 93, pp. 408--422, 2015. (journal)

\bibitem{narayanan2020first}
A. Narayanan, E. Ramadan, J. Carpenter, Q. Liu, Y. Liu, F. Qian and Z. -L. Zhang, ``A First Look at Commercial 5G Performance on Smartphones,'' {\it Proc. of the Web Conference}, pp. 894--905, 2020. (conference proceedings)

\bibitem{dimon}
R. Pueyo Centelles, M. Selimi, F. Freitag and L. Navarro, ``DIMON: Distributed Monitoring System for Decentralized Edge Clouds in Guifi.net,'' {\it Proc. of the IEEE International Conference on Service-Oriented Computing and Applications}, pp. 1--8, 2019. (conference proceedings)

\bibitem{idps-01}
M. Xiong, Y. Li, L. Gu, S. Pan, D. Zeng and P. Li, ``Reinforcement Learning Empowered IDPS for Vehicular Networks in Edge Computing,'' {\it IEEE Network}, vol. 34, no. 3, pp. 57--63, 2020. (journal)
 
\bibitem{botnet-01}
J. Ni, X. Lin and X. S. Shen, ``Toward Edge-Assisted Internet of Things: From Security and Efficiency Perspectives,'' {\it IEEE Network}, vol. 33, no. 2, pp. 50--57, 2019. (journal) 


\bibitem{electricity2030-1}
A. S. G. Andrae and T. Edler, ``On Global Electricity Usage of Communication Technology: Trends to 2030,'' {\it Challenges}, vol. 6, no. 1, pp. 117--157, 2015. (journal)

\bibitem{electricitykwhgb-1}
J. Aslan, K. Mayers, J. G. Koomey and C. France, ``Electricity Intensity of Internet Data Transmission: Untangling the Estimates,'' {\it Journal of Industrial Ecology}, vol. 22, no. 4, pp. 785--798, 2018. (journal) 

\bibitem{c02-footprint-01}
B. Ramprasad, A. Veith, M. Gabel and E. de Lara, ``Sustainable Computing on the Edge: A System Dynamics Perspective,'' {\it Proc. of the 22nd International Workshop on Mobile Computing Systems and Applications}, pp. 64--70, 2021. (workshop proceedings)


\bibitem{Yu-SP2019}
L. Yu, L. Liu, C. Pu, M. Gursoy, E. Mehmet and S. Truex, ``Differentially Private Model Publishing for Deep Learning,'' {\it Proc. of the IEEE Symposium on Security and Privacy}, pp. 332--349, 2019. (conference proceedings)

\bibitem{Gursoy-TDSC2019}
M. Gursoy, A. Tamersoy, S. Truex, W. Wei and L. Liu, ``Secure and Utility-Aware Data Collection with Condensed Local Differential Privacy,'' IEEE Transactions on Dependable and Secure Computing, 2019. (journal)

\bibitem{lockhart2020scission}
L. Lockhart, P. Harvey, P. Imai, P. Willis and B. Varghese, ``Scission: Performance-driven and Context-aware Cloud-Edge Distribution of Deep Neural Networks,'' {\it Proc. of the 13th IEEE/ACM International Conference on Utility and Cloud Computing}, pp. 257--268, 2020. (conference proceedings)


\bibitem{democratization-01}
L. Peterson, T. Anderson, S. Katti, N. McKeown, G. Parulkar, J. Rexford, M. Satyanarayanan, O. Sunay and A. Vahdat, ``Democratizing the Network Edge,'' {\it SIGCOMM Computer Communication Review}, vol. 49, no. 2, pp. 31--36, 2019. (journal)

\bibitem{globaldataplane-01}
N. Mor, R. Pratt, E. Allman, K. Lutz and J. Kubiatowicz, ``Global Data Plane: A Federated Vision for Secure Data in Edge Computing,'' {\it Proc. of the 39th IEEE International Conference on Distributed Computing Systems}, pp. 1652--1663, 2019. (conference proceedings)

\end{thebibliography}
